\documentclass[11pt]{article}

\usepackage[utf8]{inputenc}
\usepackage[english]{babel}
\usepackage[a4paper,margin=2.3cm]{geometry} %
\usepackage{graphicx} %
\usepackage{amsthm, amsmath, amssymb} 
\usepackage{tcolorbox} 
\usepackage[numbers]{natbib} %
\usepackage{lineno}	
\usepackage{appendix} %

\usepackage{authblk} 
 \usepackage{setspace} %
\usepackage{pifont}

\usepackage{hyperref}

\usepackage{placeins} %
\newcommand{\papertitle}{Less is more: removing a single bond increases the toughness of elastic networks}

\newcommand{\ie}{{\textit{i.e.}}}

\newcommand{\Gfail}{\Gamma_\mathrm{fail}}
\newcommand{\Gini}{\Gamma_\mathrm{ini}} 
\newcommand{\Gloc}{\Gamma} 

\newcommand{\Gfull}{\Gamma_\mathrm{full}}

\newcommand{\lr}{\ell_\mathrm{r}}
\newcommand{\lx}{\ell_\mathrm{x}}

\newcommand{\Lx}{L_\mathrm{x}}
\newcommand{\Ly}{L_\mathrm{y}}

\newcommand{\Lc}{L_\mathrm{c}}

\newcommand{\epsmax}{\varepsilon_\mathrm{max}}

\newcommand{\xm}{x_\mathrm{m}}
\newcommand{\ym}{y_\mathrm{m}}

\newcommand{\xt}{x_\mathrm{t}} %

\usepackage[utf8]{inputenc}
\usepackage{color}
\usepackage{xcolor}

\usepackage{amsmath}%
\usepackage{amssymb}%
\usepackage{commath}
\usepackage{csquotes}
\usepackage{graphicx}

\usepackage{svg}

\usepackage{ifthen}

\newcommand{\hidecomments}{false}

\colorlet{darkgreen}{green!50!black}

\ifthenelse{\equal{\hidecomments}{true}}{%
\newcommand{\MT}[1] {}
}{%
\newcommand{\MT}[1] {{\marginpar{\footnotesize {TRASH: #1}}}}
}

\ifthenelse{\equal{\hidecomments}{true}}{%
\newcommand{\M}[1] {}
}{%
\newcommand{\M}[1] {\colorbox{yellow}{#1}}
}

\ifthenelse{\equal{\hidecomments}{true}}{%
\newcommand{\MP}[1] {}
}{%
\newcommand{\MP}[1]{{\large \colorbox{green}{\S\S\S\S #1}}}
}

\newcommand{\A}[1]{}

\usepackage[yyyymmdd,hhmmss]{datetime}

\graphicspath{ {./} }
\begin{document}

\title{\papertitle}
\author[1]{Antoine Sanner}
\author[1]{Luca Michel} %
\author[1]{Alessandra Lingua} %

\author[1]{David S. Kammer\thanks{Corresponding Author: dkammer@ethz.ch}}

\affil[1]{Institute for Building Materials, ETH Zurich, Switzerland}

\maketitle

\section*{Abstract}

We investigate how the removal of a single bond affects the fracture behavior of triangular spring networks, whereby we systematically vary the position of the removed bond. Our simulations show that removing the bond has two contrasting effects on the fracture energy for initiation of crack propagation and on the fracture energy for failure of the entire network. A single missing bond can either lower or raise the \emph{initiation} fracture energy, depending on its placement relative to the crack tip. In contrast, the \emph{failure} fracture energy is always equal to or greater than that of a perfect network. For most initial placements of the missing bond, the crack path remains straight, and the increased failure fracture energy results from arrest at the point of maximum local fracture resistance. When the crack deviates from a straight path, we observe an even higher fracture energy, which we attribute primarily to crack bridging. This additional toughening mechanism becomes active only at low failure strains of the springs; at higher failure strains, the crack path tends to remain straight. Altogether, our results demonstrate that even a single bond removal can significantly enhance toughness, offering fundamental insights into the role of defects in polymer networks and informing the design of tough architected materials.

\newpage

\section{Introduction}  \label{sec:introduction}

Crack propagation drives the failure of brittle materials, a process that is extremely sensitive to defects such as voids or microcracks. Defects can either hinder or facilitate fracture depending on whether a dominating crack is already present or not. 
Without a preexisting crack, defects are detrimental, as they introduce stress concentrations that facilitate the initiation of a crack~\cite{alava_statistical_2006,shekhawat_damage_2013}. In contrast, when a dominating crack is already present, defects can actually be beneficial: mechanisms such as shielding of
the stress concentration~\cite{kendall_control_1975,rose_effective_1986,hutchinson_crack_1987,curtin_microcrack_1990,shum_toughening_1990,leguillon_fracture_2008,hossain_effective_2014}, deflection or meandering of the crack~\cite{urabe_fracture_2010, mirkhalaf_overcoming_2014,faber_crack_1983,faber_crack_1983a}, and crack bridging~\cite{bower_threedimensional_1991,mirkhalaf_overcoming_2014,lingua_breaking_2025}, have been shown to increase the material’s resistance to crack propagation. Thus, the presence of defects impacts the fracture of brittle materials in a non-trivial way.

Spring networks have been widely used to study the effects of defects and disorder on fracture. They provide a minimal model that can either be used as a proxy for a continuum, or to represent a variety of discrete systems such as crosslinked polymer networks~\cite{urabe_fracture_2010,arora_coarsegrained_2022,sugimura_mechanical_2013,deng_nonlocal_2023,hartquist_fracture_2024,broedersz_criticality_2011,tauber_stretchy_2022}, mechanical metamaterials~\cite{tankasala_crack_2020,waal_architecting_2024}, as well as the discrete structure of ceramics~\cite{curtin_microcrack_1990,curtin_brittle_1990}. Focusing on the propagation of a preexisting crack, two competing explanations have been proposed to explain the increase in fracture energy, the energetic cost of resisting crack propagation. \citet{urabe_fracture_2010} simulated the failure of a spring network with randomly inserted weak springs, from which they suggest that the increase in fracture energy arises from crack path roughening. While this hypothesis is qualitatively in line with other observations~\cite{lingua_breaking_2025,zhang_fiber_2017}, it remains untested and lacks a sound theoretical basis. On the other hand, \citet{curtin_microcrack_1990} attribute the toughening to elastic shielding in a similar system, whereby several springs are randomly removed. They~\cite{curtin_microcrack_1990} derive a theoretical prediction of the failure fracture energy from crack arrest statistics, which, however, has not been validated against numerical simulations, and relies on the assumption of a straight crack path. Hence, there is no consensus on the mechanism responsible for the increase in toughness due to the presence of defects. In this regard, a crucial first step is to isolate the interaction between a single defect and a propagating crack, thereby gaining a more fundamental understanding of the role of disorder in material toughening.

In this work, we investigate how a single missing bond affects the propagation of a preexisting crack in a triangular spring network. We systematically vary the position of the defect relative to the crack tip, and investigate how it influences both the initiation of crack propagation and the complete fracture of the sample.
We show that depending on its location, a single defect can either increase or decrease the fracture energy for \emph{initiation} of crack propagation, but always increases, or leaves unchanged, the fracture energy for \emph{complete failure}. Our simulations clarify the local mechanisms by which heterogeneity influences fracture resistance and lay the groundwork for understanding toughening in more complex disordered systems.

\section{Problem statement and methods} \label{sec:methods}

\begin{figure}
    \centering
    \includegraphics[width=0.6\linewidth]{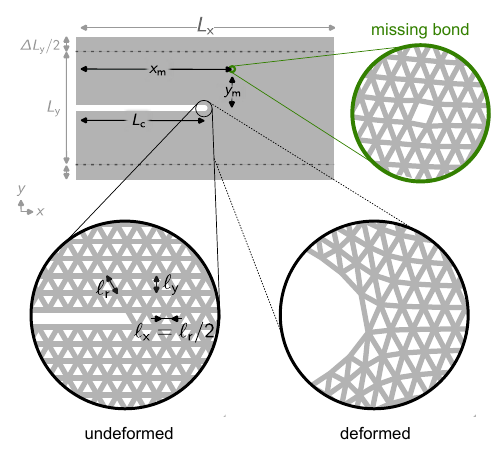}
\caption{\textbf{Pre-cracked spring network with a missing bond}. All dimensions are given in the undeformed state of the network, where all springs are at their rest length $\lr$. The deformed configuration is shown at the onset of crack propagation, when the spring at the crack tip reaches its maximum length $ (1 + \epsmax) \lr$ with $\epsmax = 1$.}
    \label{fig:setup}
\end{figure}

We consider the failure of a triangular network of linear springs under quasi-static conditions. In the undeformed state, all springs have a rest length $\lr$ and carry no force. Upon deformation, the force in a spring is $f = k(\ell - \lr)$, where $\ell$ is the deformed length and $k$ is the spring stiffness. A spring breaks abruptly when its strain $\varepsilon = \ell / \lr - 1$ exceeds a critical threshold $\epsmax = 1$. 

We use a thin-strip geometry~\cite{deng_nonlocal_2023,rivlin_rupture_1953,kermode_lowspeed_2008,long_fracture_2016} with a width $\Lx = 200 \lr$, height $\Ly = 100 \sqrt{3}/2 \lr$ (equal to the height of 100 unit cells), and a pre-crack of length $\Lc = 100 \lr$ (see Fig.~\ref{fig:setup}). These dimensions are large enough to minimize boundary effects. We apply a uniform vertical displacement on the top and bottom boundaries and constrain lateral displacements to zero on all boundaries.

We load the system by applying increments of vertical displacement $\Delta L_y$ until the crack has propagated through the entire network. At each increment, we solve for static equilibrium using the FIRE minimization algorithm implemented in LAMMPS~\cite{bitzek_structural_2006,guenole_assessment_2020,thompson_lammps_2022}. If any spring exceeds the critical strain $\epsmax$, we remove the most stretched one, following the approach by~\citet{dussi_athermal_2020}. We then resolve the equilibrium and repeat this cycle until no further springs break. After that, we apply the next displacement increment.

We quantify the resistance of the network to crack propagation by the fracture energy $\Gamma$, defined as the critical elastic energy release rate $G$ required to propagate the crack. We compute $G$ as the elastic energy density in the region $x \in [3L_x/4, L_x]$ ahead of the crack tip, multiplied by the strip height $L_y$. This measure is independent of the crack length and becomes exact in the limit of an infinitely long strip ($L_c / L_y \to \infty$, $L_x / L_y \to \infty$) due to translational symmetry. Hence, in this setup, applying vertical displacements is equivalent to prescribing an energy release rate.

\section{Results} \label{sec:results}

\FloatBarrier

\begin{figure}
    \centering
    \includegraphics[width=0.6\linewidth]{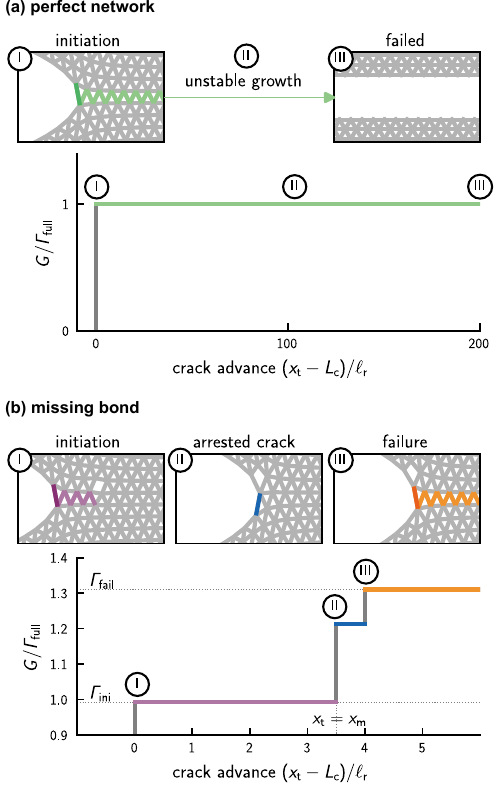}
    \caption{\textbf{Removing a bond from a perfect network increases the fracture energy due to crack arrest}. Applied elastic energy release rate $G$ as a function of crack advance for (a) a pre-cracked but otherwise perfect network, and (b) a pre-cracked network with a missing bond. $G$ is normalized by the fracture energy of the perfect network $\Gfull$. Colored segments of the curves indicate unstable crack growth events. (insets a-I and b-I to III) Deformed configurations at the onset of the unstable event corresponding to each label. Bonds broken during the unstable event are highlighted in color, with darker colors indicating those that break first. (inset a-III) Final configuration of the failed network after the unstable event.
    }
    \label{fig:Gini-Gbreak-rcurves}
\end{figure}

\subsection{Reference case: perfect network} \label{sec:perfect}

As a reference case, we first consider a pre-cracked but otherwise perfect triangular network and load it until failure. In this case, the first bond to break is the one at the crack tip (see Fig.~\ref{fig:Gini-Gbreak-rcurves}a-I). Its failure triggers the unstable breaking of all remaining bonds along the crack plane. As a result, the network fails instantaneously along a perfectly straight crack path (see Fig.~\ref{fig:Gini-Gbreak-rcurves}a-III). This sudden failure occurs at a constant energy release rate (see Fig.~\ref{fig:Gini-Gbreak-rcurves}a), which is the driving force for crack propagation. The fracture resistance is therefore characterized by a single value: the critical energy release rate at initiation of propagation, which we denote as the fracture energy $\Gfull$. This value serves as a reference for the imperfect networks studied below.

\subsection{Removing a single bond: effect on crack propagation}

We now consider the failure of networks with a single missing bond, located at position $\xm, \ym$ (see Fig.~\ref{fig:setup}). In the specific case considered here, the missing bond lies one lattice plane above the crack plane and three unit cells ahead of the crack tip (see Fig.~\ref{fig:Gini-Gbreak-rcurves}b-I). 

The missing bond changes the crack propagation compared to the perfect network. Similarly to before, the first bond to break is the one at the crack tip. However, this first bond failure now occurs at an applied energy release rate $G$ that is slightly lower than the reference value $\Gfull$ of the perfect network. The breaking of the first bond is followed by unstable crack growth, similar to the perfect network, but it does not lead to complete failure. Instead, the crack advances by breaking only seven bonds (see Fig.~\ref{fig:Gini-Gbreak-rcurves}b-I) and arrests just below the missing bond, \textit{i.e.}, when crack tip position $\xt = \xm$ (see Fig.~\ref{fig:Gini-Gbreak-rcurves}b-II). The crack remains pinned at this location while $G$ continues to increase with further applied displacement. It then advances by one bond and arrests again (see Fig.~\ref{fig:Gini-Gbreak-rcurves}b-II). After a further increase in $G$, the crack eventually propagates catastrophically across the entire strip (see Fig.~\ref{fig:Gini-Gbreak-rcurves}b-III). Overall, these results show that the missing bond acts as an obstacle to crack propagation, leading to an increased energy release rate required for the network to fail completely.

Since $G$ increases between the initiation of crack growth and final failure, the resistance to crack propagation can no longer be characterized by a single fracture energy, unlike for the perfect network. We therefore define $\Gini$ as the energy release rate at initiation, and $\Gfail$ as the energy release rate required for complete failure. While for a perfect network this means $\Gfull = \Gini = \Gfail$, the missing bond in Fig.~\ref{fig:Gini-Gbreak-rcurves}b leads to $\Gfail > \Gini$. More importantly, while the first bond breaks at $\Gini < \Gfull$ due to the missing bond, complete failure occurs only at $\Gfail > \Gfull$, indicating that, at least for the case in Fig.~\ref{fig:Gini-Gbreak-rcurves}b, removing a bond increases the toughness despite facilitating initiation. 

\begin{figure}
    \centering
    \includegraphics[width=0.9\linewidth]{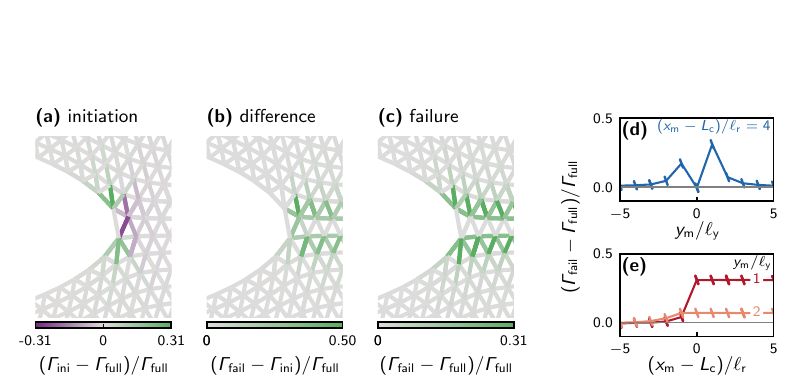}
    \caption{\textbf{A missing bond increases the fracture energy for failure, regardless of its position.} Dependence of (a) the initiation fracture energy $\Gini$, (b) the difference of the failure fracture energy $\Gfail$ to $\Gini$, and (c) the failure fracture energy $\Gfail$ on the position of the missing bond. The color of each bond represents how its removal affects the fracture energy. (d, e) Dependence of $\Gfail$ on the position of the missing bond when either its $y$-coordinate $\ym$ (d) or $x$-coordinate $\xm$ (e) is fixed. Only bonds oriented toward the upper left, as indicated by the marker shape, are shown.
    }
    \label{fig:GiniGbreak-maps}
\end{figure}

\subsection{Systematic variation of missing bond placement}

To test the generality of the observations made in the previous section, we now investigate how missing bonds affect the initiation and failure fracture energies for all possible missing bond locations. To this end, we systematically remove a single bond and compare the resulting $\Gini$ and $\Gfail$ of the disrupted network to $\Gfull$, the reference value of the perfect network. 

We focus first on crack initiation, characterized by the initiation fracture energy $\Gini$. We observe that $\Gini$ varies with the position of the missing bond, as shown by the color coding in Fig.~\ref{fig:GiniGbreak-maps}a. More specifically, missing bonds located ahead of the crack tip tend to reduce $\Gini$ relative to $\Gfull$. In contrast, missing bonds behind the crack tip increase $\Gini$. The influence of bond removal becomes stronger the closer the bond is to the crack tip. Overall, removing a bond behind the crack tip hinders crack growth initiation, while removing one ahead facilitates it. These observations are consistent with earlier studies on pairs of missing bonds in triangular spring networks~\cite{curtin_brittle_1990} and on microcracks in continuum materials~\cite{shum_toughening_1990}.

Next, we consider the failure fracture energy $\Gfail$. When the missing bond is located behind the crack tip, we observe $\Gfail = \Gini$ (see Fig.~\ref{fig:GiniGbreak-maps}b), which is the signature of abrupt failure, as seen in the perfect network (see Fig.~\ref{fig:Gini-Gbreak-rcurves}a). In contrast, missing bonds ahead of the crack tip lead to $\Gfail > \Gini$, meaning that the resistance to failure increases during crack propagation, as in the example shown in Fig.~\ref{fig:Gini-Gbreak-rcurves}b. While $\Gfail \geq \Gini$ is expected because crack initiation must precede failure, it is more remarkable that, compared to the perfect network, $\Gfail \geq \Gfull$, regardless of the position of the missing bond (see Fig.~\ref{fig:GiniGbreak-maps}c). In other words, compared to a perfect network, removing a bond may increase, but never decrease, the failure fracture energy $\Gfail$.

Finally, we note that missing bonds located closer to the crack plane have a more pronounced effect on $\Gfail$ (see Fig.~\ref{fig:GiniGbreak-maps}d), except in the plane itself, where $\Gfail = \Gfull$. More interestingly, ahead of the crack tip, $\Gfail$ is independent of the $x$-position of the missing bond (see Fig.~\ref{fig:GiniGbreak-maps}e). While the sharp dependence on $\ym$ is consistent with the high sensitivity of $\Gini$ near the crack tip. The extended influence for $\xm > \xt$ is, as will be discussed below, a direct consequence of crack growth.

\section{Mechanism of toughening by a missing bond: crack arrest by local perturbations} \label{sec:mec}

Our numerical results demonstrate that removing a single bond in a triangular lattice affects both the initiation fracture energy $\Gini$ and the failure fracture energy $\Gfail$. While $\Gfail$ is always greater than $\Gfull$, $\Gini$ may be either smaller or larger than $\Gfull$, depending on the position of the missing bond. We examine the mechanism responsible for $\Gfail$ being larger than $\Gini$, focusing on the simplified case of a straight crack path. This assumption allows us to apply a crack arrest argument~\cite{curtin_microcrack_1990,charles_crack_2002} that links $\Gfail$ to the sequence of local initiation energies $\Gloc$ encountered during crack propagation. As a result, understanding how the missing bond perturbs $\Gini$ is key to understanding the observed increase in $\Gfail$. We begin by investigating how bond removal alters the force distribution in the network, and how these local force perturbations translate into changes in $\Gini$. The validity of the straight crack path assumption will be addressed in Section~\ref{sec:validity}.

\subsection{The link between $\Gini$ and local force perturbations} \label{sec:mec:Gini}

We begin by examining how changes in $\Gini$ arise from changes in force distributions near the crack tip. In the perfect case, without a missing bond, the forces are concentrated near the crack tip, with the highest force in the bond immediately ahead of the crack tip, labeled $t$ (see the \textit{full} system in Fig.~\ref{fig:stress-perturb}a). As the network is loaded, bond $t$ is the first to fail, and its force, $f_t$, determines $\Gini$. Hence, a perturbation of $f_t$, such as caused by a removed bond, will directly affect crack growth initiation and $\Gini$. Specifically, if the removal of a bond reduces $f_t$, a higher applied $G$ is required to break bond $t$, thus increasing $\Gini$. Conversely, if $f_t$ increases, $\Gini$ is reduced.

The removal of a bond redistributes the forces in the neighboring bonds, causing changes in forces compared to the perfect network, as shown in Fig.~\ref{fig:stress-perturb}b. This effect is localized with force perturbations decreasing rapidly within a few layers of distance to the missing bond. Consequently, when the missing bond is far from the crack tip, as in case A, the force in the critical bond $f_t$ is largely unaffected (see Fig.~\ref{fig:stress-perturb}b), and hence $\Gini$ is essentially equal to that of the full network.

Another characteristic of the force perturbation is that some bonds experience an increase in force whereas others a decrease (see Fig.~\ref{fig:stress-perturb}b case A). This explains why $\Gini$ may decrease or increase depending on the position of the removed bond. For instance, when the missing bond is just ahead of the crack tip, in position B, bond $t$ is among those experiencing an increased force. Consequently, the force in bond $t$ is elevated compared to the full case, as shown in Fig.~\ref{fig:GiniGbreak-maps}b, leading to $\Gini < \Gfull$. Conversely, when the missing bond is just above the crack tip, in position C, bond $t$ lies in a region of reduced force, resulting in $\Gini > \Gfull$.

Overall, these results show that the force $f_t$ at the crack tip may be increased or decreased depending on the position of the removed bond. These variations in $f_t$ allow for rationalizing the observed decrease and increase in $\Gini$ compared to the perfect network.

\subsection{Evolution of fracture energy during forced crack growth} \label{sec:mec:growth}

Looking beyond the initiation of crack growth, we now focus on how a missing bond affects the resistance to crack growth as the crack is approaching the missing bond, passing it, and eventually leaving it behind (see Fig.~\ref{fig:advancing-ini}a). For simplicity, we will focus on cases where the crack path is straight, and take as an example the case where the initial position of the missing bond is in position A, as depicted in Fig.~\ref{fig:advancing-ini}a-A.

To evaluate the effect of the missing bond on the growing crack, we first consider the case of a forced crack growth at a fixed applied displacement, \ie, the crack length is artificially increased as opposed to spontaneous propagation considered later. In fact, configurations with a missing bond at a fixed position with coordinates $\xm, \ym$ and an evolving crack tip position are equivalent to the case of crack initiation of missing bonds at varying positions. This is because the thin-strip geometry is nearly translationally invariant, so only the relative distance $\xt - \xm$ matters. 

\begin{figure}
    \centering
    \includegraphics[width=0.9\linewidth]{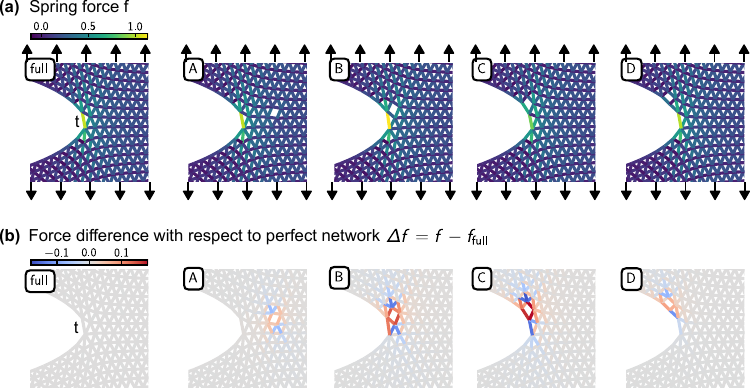}
    \caption{
    \textbf{A missing bond increases or decreases the force in the bond at the crack tip depending on its position.} (a) Bond forces due to the applied load for a perfect pre-cracked network labeled full and for different positions of the missing bond labeled A to D. (b) Difference in bond forces in the network with a missing bond relative to the network without a missing bond.
    }
    \label{fig:stress-perturb}
\end{figure}

\begin{figure}
    \centering
    \includegraphics[width=0.65\linewidth]{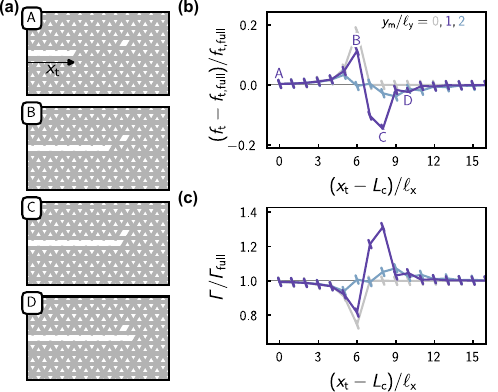}
    \caption{\textbf{Local fracture energy $\Gloc$ always decreases and then increases with crack advance.}
    (a) Advancing crack with fixed absolute positions of the missing bond. The configurations A to D are equivalent to the configurations A to D shown in Fig.~\ref{fig:stress-perturb}b due to the translational symmetry of the boundary conditions. (b) Evolution of force in the bond at the crack tip and (c) local fracture energy $\Gloc$ as a function of crack advance, measured in terms of number of broken bonds.}
    \label{fig:advancing-ini}
\end{figure}

Consequently, we can consider crack advance as A-B-C-D being a sequence of configurations visited during crack propagation. This means that by considering Fig.~\ref{fig:stress-perturb}b, we observe that at the start of propagation (position A), $f_t$ is nearly equal to $f_\mathrm{t,full}$ because the crack tip is still far from the missing bond, and the force perturbation decays rapidly with distance. As the crack approaches the missing bond, $f_t$ increases (see Fig.~\ref{fig:advancing-ini}b), entering a region where the missing bond significantly amplifies local forces. The highest $f_t$ occurs at position B, where the missing bond is one unit cell ahead of the crack tip. As soon as the crack passes this point, $f_t$ decreases sharply and becomes lower than $f_\mathrm{t,full}$. The lowest value is reached at position C, after two additional bonds have broken. Once the crack advances further, beyond the missing bond (position D), $f_t$ increases again but remains slightly below $f_\mathrm{t,full}$. This overall behavior is independent of the $y$ position and orientation of the missing bond (see Fig.~\ref{fig:advancing-ini}b and Appendix~\ref{app:positiondependence}).

Next, we consider the critical energy release required for crack advance, at any given position, which we denote as the local fracture energy $\Gloc$. We determine $\Gloc$ by increasing the displacement applied to the network until the bond at the crack tip reaches the critical strain $\epsmax=1$. The elastic energy release rate corresponding to the critical applied displacement yields $\Gloc$. Since an increase in the force of the critical bond decreases the required energy release rate to advance the crack, the evolution of $\Gloc$ as the crack advances takes the same shape mirrored around the horizontal axis (compare Figs.~\ref{fig:advancing-ini}b and c). Hence, any missing bond lying ahead of the crack tip will always first facilitate crack propagation but later act as an obstacle once the crack passes it. As for the force perturbation, this overall behavior is independent of the $y$ position and orientation of the missing bond.

\subsection{Crack arrest during spontaneous propagation} %
\label{sec:straight-theory}

After determining the local fracture energy for a forced propagation, we now focus on spontaneous crack growth under increasing applied energy release rates. We explain how the applied energy release rate $G$ relates to the local fracture energy $\Gloc$ based on the principle that the crack can only advance if the applied $G$ is greater than or equal to the local $\Gloc$. 

\begin{figure}
    \centering
 \includegraphics[width=0.6\linewidth]{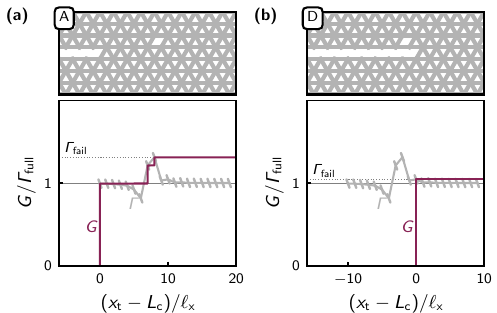}
    \caption{\textbf{Origin of the crack propagation behavior based on the local fracture energy.}
    Top panels: initial relative position of the crack and the missing bond. Bottom panels: Applied energy release rate as a function of crack tip position in a crack propagation for cases where the missing bond is (a) ahead of, and (b) behind the crack tip. The continuous red line is the applied energy release rate in the crack propagation simulation, and the gray line with oblique markers is the local fracture energy $\Gloc$.
    }
    \label{fig:single_bond_propagation}
\end{figure}

As the applied energy release rate $G$ increases, the crack begins to propagate once $G$ reaches the local fracture energy at initial crack length $\Gini = \Gloc (\Lc)$ (see Fig.~\ref{fig:single_bond_propagation}a). After the first bond broke, the local fracture energy $\Gloc (\Lc + \lx)$ is slightly lower than $\Gloc(\Lc)$, and still less than $G$; hence, the crack continues to advance. This process repeats for the next $n$ bonds, as long as $\Gloc(\Lc + n \lx) < G$, the crack tip position being $\xt = \Lc + n \lx$. However, when the crack reaches a point where $\Gloc$ suddenly exceeds $G$, specifically $n=7$ for the case shown in Fig.~\ref{fig:single_bond_propagation}a, the crack arrests. At this point, further propagation requires an increase in $G$ to match the fracture energy at this location.

The crack will only propagate through the entire strip once $G$ exceeds the maximum local fracture energy encountered along its path. Thus, the failure fracture energy $\Gfail$ corresponds to the largest value of $\Gloc$ during propagation: 
\begin{equation} \label{eq:GfailfromGini} 
\Gfail = \underset{0 \leq n}{\max} ~\Gloc(\Lc + n \lx)~. 
\end{equation}
Hence, the crack is arrested by the strongest obstacle along its path, \ie, the position with the highest $\Gloc$, explaining both the increasing $G$ during propagation and why $\Gfail \geq \Gini$.

We now generalize the above result by considering how the position of the missing bond relative to the initial crack tip affects the shape of the R-curve and the resulting failure energy. We have shown that $\Gamma(\xt)$ always has the same shape, first decreasing with $\xt$, abruptly increasing above $\Gfull$, followed by a decay approaching $\Gfull$ from above. There are two cases to consider:

(i) When the missing bond is initially ahead of the crack tip, as in the example discussed above and shown in Fig.~\ref{fig:single_bond_propagation}a, the crack tip eventually reaches the location of the defect, where $\Gloc$ reaches its maximum. Consequently, $\Gfail$ equals this peak value, regardless of the initial distance between the crack tip and the missing bond. This explains why $\Gfail$ is independent of the bond’s $x$-position when $\xt - \xm < 0$, as shown in Fig.~\ref{fig:GiniGbreak-maps}e.

(ii) When the missing bond is initially behind the crack tip, as in Fig.~\ref{fig:single_bond_propagation}b, the maximum of $\Gloc$ lies behind the crack front. Since the crack does not need to overcome this obstacle, $\Gfail$ is smaller than in the case where the bond is ahead. Moreover, because $\Gloc$ typically decreases with crack advance, failure occurs abruptly, and $\Gfail = \Gini$, as shown in Fig.\ref{fig:GiniGbreak-maps}c.

Hence, the final failure always occurs in a region where $\Gamma(\xt) \geq \Gfull$, \ie{}, either when the crack tip is at the peak in $\Gamma$ or beyond it. This means that $\Gfail > \Gfull$ as observed in Fig.~\ref{fig:GiniGbreak-maps}c for all placements of the missing bond except when it is within the crack plane. In that case, the missing bond gets absorbed by the crack tip before it can increase $\Gamma$ and $\Gfail = \Gfull$. In summary, the presence of a missing bond never reduces, and often increases, the fracture energy for complete failure $\Gfail$ compared to that of a perfect network.

\subsection{Validity of the assumption of a straight crack path}
\label{sec:validity}

\begin{figure}
    \centering
    \includegraphics[width=0.65\linewidth]{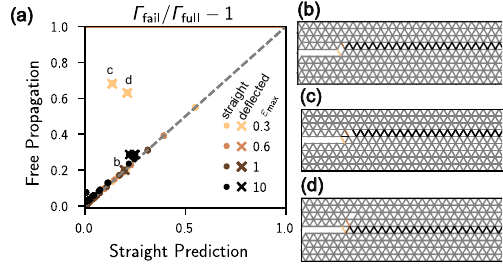}
    \caption{
\textbf{Assuming a straight crack path yields a lower bound for the fracture energy.}
(a) Comparison between the failure fracture energy $\Gfail$ obtained from simulations and the prediction from Eq.~\ref{eq:GfailfromGini}, for various values of the critical bond strain $\epsmax$. Each data point corresponds to a different position of the missing bond. Simulations in which the crack propagated along a straight path are shown as dots; simulations where the crack deviated from the original crack plane are marked with crosses.
(b) Example of a non-straight crack path for $\epsmax = 1$. Unbroken bonds are shown in grey. Broken bonds are color-coded by the order in which they broke: earlier events appear brighter, and the final failure event is shown in black. Note that all bond positions are shown in the undeformed (reference) configuration of the network.
(c,d) Crack paths for two defect placements that lead to the largest increase in $\Gfail$ relative to the straight-path prediction for $\epsmax = 0.3$.}
    \label{fig:validation_all_fig7}
\end{figure}

The prediction of $\Gfail$ given by Eq.~\ref{eq:GfailfromGini} assumes that the crack propagates strictly along a straight path. This assumption holds in many cases, including those shown in Figs.~\ref{fig:advancing-ini} and \ref{fig:single_bond_propagation}, but not for all possible defect positions. To test the validity of this assumption, we compare the predicted $\Gfail$ to results from simulations in which the crack is free to deviate from a straight path (Fig.~\ref{fig:validation_all_fig7}), while systematically varying the position of the missing bond. In addition to the spring failure strain $\epsmax = 1$ considered so far, we extend this comparison to the range $\epsmax = 0.3$ to $10$. Most defect placements result in straight crack paths, but some lead to bond breaking away from the initial crack plane, such as the example shown in Fig.~\ref{fig:validation_all_fig7}b,c and d. The simulations where the crack deviated from a straight path are marked with crosses in Fig.~\ref{fig:validation_all_fig7}a. In all such cases, the measured $\Gfail$ is greater than or equal to the straight-path prediction. We conclude that Eq.~\ref{eq:GfailfromGini} provides a lower bound for $\Gfail$ in systems where the crack is free to deviate from a straight path.

\section{Discussion}

In this study, we examined how the removal of a single bond affects the fracture toughness of triangular spring networks, using both simulations and theoretical analysis. Our central finding is that removing just one bond never reduces, and often increases, the fracture energy for complete failure, in line with previous observations on systems with small concentrations of defects~\cite{abid_fracture_2019,deng_nonlocal_2023,hartquist_fracture_2024,hossain_effective_2014}. In most cases, the crack path remains straight, allowing the increase in fracture energy to be predicted using classical crack arrest theories developed for planar fracture~\cite{curtin_brittle_1990,charles_crack_2002}. When the crack deviates from a straight path, the measured $\Gfail$ exceeds the prediction, demonstrating that Eq.~\ref{eq:GfailfromGini} provides a robust lower bound.

Remarkably, up to a 50\% increase in toughness can be achieved without any deflection of the crack path. This demonstrates that crack path roughening is not necessary to enhance fracture energy in spring networks. However, our results also support the idea that roughening contributes to toughening~\cite{urabe_fracture_2010,zhang_fiber_2017}: in cases where the crack does deviate, the resulting $\Gfail$ is even higher than in the straight-path cases. Hence, crack arrest by local heterogeneity and crack deflection are both important toughening mechanisms.

A possible explanation for toughness increase via crack roughening is the increased number of broken springs associated with a longer, more tortuous crack path~\cite{zhang_fiber_2017,persson_effect_2001}. However, in our simulations, the crack typically deflects only once and remains straight thereafter (see Fig.~\ref{fig:validation_all_fig7}a–c). Consequently, the total number of broken bonds increases only marginally and cannot explain the up to threefold increase in fracture energy compared to the straight-path prediction. Our results indicate that resistance to crack propagation is governed not by average quantities such as the number of broken bonds along the entire crack path, but by the most severe local obstacle encountered along the crack path.

Additional toughening mechanisms likely contribute in cases with deflected cracks. For instance, at low $\epsmax = 0.3$, we observe crack bridging (Fig.~\ref{fig:validation_all_fig7}c), where a spring spans the crack faces before breaking, as observed in lattices of stiff trusses~\cite{lingua_breaking_2025} and similar to bridging zones in composite materials. Other examples show off-plane bond failure ahead of the crack tip (Fig.~\ref{fig:validation_all_fig7}d), reminiscent of microcracking or transformation-induced toughening~\cite{evans_toughening_1981,kreher_increased_1981,evans_toughening_1986}. While these effects are weak at $\epsmax = 1$, they become more prominent as $\epsmax$ decreases and the crack path becomes increasingly tortuous. The increasing path complexity and fracture energy at low $\epsmax$ suggest greater sensitivity to disorder and finite-size effects in this regime. For $\epsmax < 0.3$, our brute-force simulations fail to converge, indicating that much larger systems, and possibly multiscale numerical methods~\cite{deng_nonlocal_2023,pastewka_seamless_2012,ma_hybrid_2019,sinclair_influence_1975}, are needed to accurately probe this limit.

This trend toward more tortuous crack paths at low $\epsmax$ may also generalize to architected materials. Specifically, it suggests that lattices made from highly stretchable trusses, such as those used in Refs.\cite{deng_nonlocal_2023,hartquist_scaling_2025,hartquist_fracture_2024}, could exhibit straighter crack paths than conventional lattices of stiff trusses, which often fracture along highly tortuous paths~\cite{lingua_breaking_2025}.

\section{Conclusion}

We have investigated how the removal of a single bond in a triangular spring network affects the propagation of a preexisting crack. Focusing first on the initiation of crack propagation, we showed that the initiation fracture energy $\Gini$ is sensitive to the position of the missing bond. Removing a bond ahead of the crack tip increases the force concentration at the tip and lowers $\Gini$, while removing a bond above or below the crack tip reduces this concentration, thereby increasing $\Gini$.
In contrast, the failure fracture energy $\Gfail$ is always greater than or equal to that of a perfect network. This is because a missing bond that initially lowers $\Gini$ eventually ends up in a position where it impedes further crack propagation. Thus, even a single, isolated defect never reduces the fracture energy; instead, it often increases it.

\section{Acknowledgements}
The authors acknowledge Jan van Dokkum and Mohit Pundir for useful discussions. The authors acknowledge the Swiss National Science Foundation for financial support under grant number 200021\_200343.

\section{CRediT authorship contribution statement}
\textbf{Antoine Sanner}: Conceptualization, Methodology, Investigation, Formal Analysis, Data Curation and Visualisation, Writing -- Original Draft.\\
\textbf{Luca Michel}: Conceptualization, Writing -- Review and Editing\\
\textbf{Alessandra Lingua}: Writing -- Review and Editing\\
\textbf{David S. Kammer}: Conceptualisation, Supervision, Formal Analysis, Writing -- Review and Editing, Project administration, Funding acquisition.

\section{Declaration of competing interest}
The authors declare that they have no known competing financial interests or personal relationships that could have appeared to influence the work reported in this paper.

\section{Code Availability}
The code used for the numerical simulations is available on ETH GitLab: [URL].

\section{Data Availability}
The data that support the findings of this study are openly available in ETH Research Collection at [URL].

\appendix

\section{Dependence of the crack tip force on the position of the missing bond} \label{app:positiondependence}

\begin{figure}
    \centering
    \includegraphics{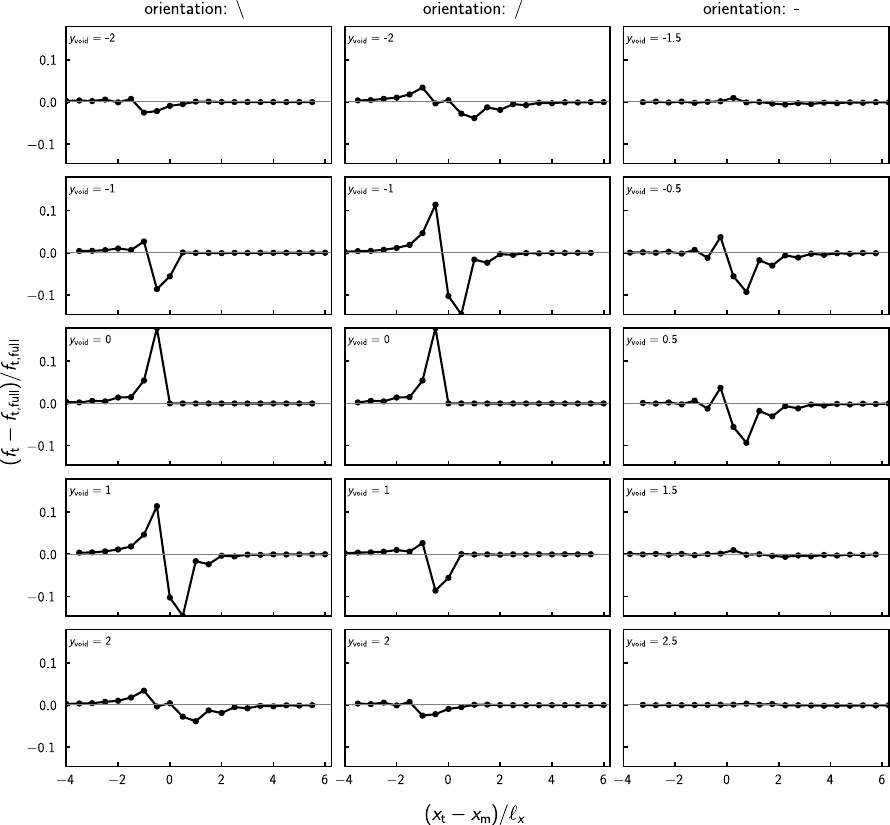}
    \caption{
Evolution of the force in the bond at the crack tip as a function of crack advance, showing that the force always has a peak followed by a dip and finally converges back to the perfect network value, irrespective of the $y$-position of the removed bond $\ym$ and its orientation. Here we report the crack tip position $\xt$ with respect to the position of the missing bond $\xm$. 
    }
    \label{fig:sup:force_updown_all}
\end{figure}

The observed sequence of a peak of $f_t$ followed by a decrease below $f_\mathrm{t,full}$ is independent of the orientation of the missing bond, as shown in Fig.~\ref{fig:sup:force_updown_all}. 

\begin{thebibliography}{42}
\providecommand{\natexlab}[1]{#1}
\providecommand{\url}[1]{\texttt{#1}}
\expandafter\ifx\csname urlstyle\endcsname\relax
  \providecommand{\doi}[1]{doi: #1}\else
  \providecommand{\doi}{doi: \begingroup \urlstyle{rm}\Url}\fi

\bibitem[Alava et~al.(2006)Alava, Nukala, and Zapperi]{alava_statistical_2006}
Mikko~J. Alava, Phani K. V.~V. Nukala, and Stefano Zapperi.
\newblock Statistical models of fracture.
\newblock \emph{Advances in Physics}, 55\penalty0 (3-4):\penalty0 349--476, May
  2006.
\newblock ISSN 0001-8732.
\newblock \doi{10.1080/00018730300741518}.

\bibitem[Shekhawat et~al.(2013)Shekhawat, Zapperi, and
  Sethna]{shekhawat_damage_2013}
Ashivni Shekhawat, Stefano Zapperi, and James~P. Sethna.
\newblock From {{Damage Percolation}} to {{Crack Nucleation Through Finite Size
  Criticality}}.
\newblock \emph{Phys. Rev. Lett.}, 110\penalty0 (18):\penalty0 185505, April
  2013.
\newblock \doi{10.1103/PhysRevLett.110.185505}.

\bibitem[Kendall and Cottrell(1975)]{kendall_control_1975}
K.~Kendall and Alan~Howard Cottrell.
\newblock Control of cracks by interfaces in composites.
\newblock \emph{Proceedings of the Royal Society of London. A. Mathematical and
  Physical Sciences}, 341\penalty0 (1627):\penalty0 409--428, January 1975.
\newblock \doi{10/bn3htd}.

\bibitem[Rose(1986)]{rose_effective_1986}
L.~R.~F. Rose.
\newblock Effective {{Fracture Toughness}} of {{Microcracked Materials}}.
\newblock \emph{Journal of the American Ceramic Society}, 69\penalty0
  (3):\penalty0 212--214, 1986.
\newblock ISSN 1551-2916.
\newblock \doi{10.1111/j.1151-2916.1986.tb07409.x}.

\bibitem[Hutchinson(1987)]{hutchinson_crack_1987}
J.~W. Hutchinson.
\newblock Crack tip shielding by micro-cracking in brittle solids.
\newblock \emph{Acta Metallurgica}, 35\penalty0 (7):\penalty0 1605--1619, July
  1987.
\newblock ISSN 0001-6160.
\newblock \doi{10.1016/0001-6160(87)90108-8}.

\bibitem[Curtin and Futamura(1990)]{curtin_microcrack_1990}
W.~A. Curtin and K.~Futamura.
\newblock Microcrack toughening?
\newblock \emph{Acta Metallurgica et Materialia}, 38\penalty0 (11):\penalty0
  2051--2058, November 1990.
\newblock ISSN 0956-7151.
\newblock \doi{10.1016/0956-7151(90)90072-O}.

\bibitem[Shum and Hutchinson(1990)]{shum_toughening_1990}
David K.~M. Shum and John~W. Hutchinson.
\newblock On toughening by microcracks.
\newblock \emph{Mechanics of Materials}, 9\penalty0 (2):\penalty0 83--91,
  September 1990.
\newblock ISSN 0167-6636.
\newblock \doi{10.1016/0167-6636(90)90032-B}.

\bibitem[Leguillon and Piat(2008)]{leguillon_fracture_2008}
D.~Leguillon and R.~Piat.
\newblock Fracture of porous materials -- {{Influence}} of the pore size.
\newblock \emph{Engineering Fracture Mechanics}, 75\penalty0 (7):\penalty0
  1840--1853, May 2008.
\newblock ISSN 0013-7944.
\newblock \doi{10.1016/j.engfracmech.2006.12.002}.

\bibitem[Hossain et~al.(2014)Hossain, Hsueh, Bourdin, and
  Bhattacharya]{hossain_effective_2014}
M.~Z. Hossain, C.~J. Hsueh, B.~Bourdin, and K.~Bhattacharya.
\newblock Effective toughness of heterogeneous media.
\newblock \emph{Journal of the Mechanics and Physics of Solids}, 71:\penalty0
  15--32, November 2014.
\newblock ISSN 0022-5096.
\newblock \doi{10.1016/j.jmps.2014.06.002}.

\bibitem[Urabe and Takesue(2010)]{urabe_fracture_2010}
Chiyori Urabe and Shinji Takesue.
\newblock Fracture toughness and maximum stress in a disordered lattice system.
\newblock \emph{Phys. Rev. E}, 82\penalty0 (1):\penalty0 016106, July 2010.
\newblock \doi{10.1103/PhysRevE.82.016106}.

\bibitem[Mirkhalaf et~al.(2014)Mirkhalaf, Dastjerdi, and
  Barthelat]{mirkhalaf_overcoming_2014}
M.~Mirkhalaf, A.~Khayer Dastjerdi, and F.~Barthelat.
\newblock Overcoming the brittleness of glass through bio-inspiration and
  micro-architecture.
\newblock \emph{Nat Commun}, 5\penalty0 (1):\penalty0 3166, January 2014.
\newblock ISSN 2041-1723.
\newblock \doi{10.1038/ncomms4166}.

\bibitem[Faber and Evans(1983{\natexlab{a}})]{faber_crack_1983}
K.~T. Faber and A.~G. Evans.
\newblock Crack deflection processes---{{I}}. {{Theory}}.
\newblock \emph{Acta Metallurgica}, 31\penalty0 (4):\penalty0 565--576, April
  1983{\natexlab{a}}.
\newblock ISSN 0001-6160.
\newblock \doi{10.1016/0001-6160(83)90046-9}.

\bibitem[Faber and Evans(1983{\natexlab{b}})]{faber_crack_1983a}
K.~T. Faber and A.~G. Evans.
\newblock Crack deflection processes---{{II}}. {{Experiment}}.
\newblock \emph{Acta Metallurgica}, 31\penalty0 (4):\penalty0 577--584, April
  1983{\natexlab{b}}.
\newblock ISSN 0001-6160.
\newblock \doi{10.1016/0001-6160(83)90047-0}.

\bibitem[Bower and Ortiz(1991)]{bower_threedimensional_1991}
A.~F. Bower and M.~Ortiz.
\newblock A three-dimensional analysis of crack trapping and bridging by tough
  particles.
\newblock \emph{Journal of the Mechanics and Physics of Solids}, 39\penalty0
  (6):\penalty0 815--858, January 1991.
\newblock ISSN 0022-5096.
\newblock \doi{10.1016/0022-5096(91)90026-K}.

\bibitem[Lingua et~al.(2025)Lingua, Sanner, Hild, and
  Kammer]{lingua_breaking_2025}
Alessandra Lingua, Antoine Sanner, Fran{\c c}ois Hild, and David~S. Kammer.
\newblock Breaking better: {{Imperfections}} increase fracture resistance in
  architected lattices, April 2025.

\bibitem[Arora et~al.(2022)Arora, Lin, and Olsen]{arora_coarsegrained_2022}
Akash Arora, Tzyy-Shyang Lin, and Bradley~D. Olsen.
\newblock Coarse-{{Grained Simulations}} for {{Fracture}} of {{Polymer
  Networks}}: {{Stress Versus Topological Inhomogeneities}}.
\newblock \emph{Macromolecules}, 55\penalty0 (1):\penalty0 4--14, January 2022.
\newblock ISSN 0024-9297.
\newblock \doi{10.1021/acs.macromol.1c01689}.

\bibitem[Sugimura et~al.(2013)Sugimura, Asai, Matsunaga, Akagi, Sakai, Noguchi,
  and Shibayama]{sugimura_mechanical_2013}
Asumi Sugimura, Makoto Asai, Takuro Matsunaga, Yuki Akagi, Takamasa Sakai,
  Hiroshi Noguchi, and Mitsuhiro Shibayama.
\newblock Mechanical properties of a polymer network of {{Tetra-PEG}} gel.
\newblock \emph{Polym J}, 45\penalty0 (3):\penalty0 300--306, March 2013.
\newblock ISSN 1349-0540.
\newblock \doi{10.1038/pj.2012.149}.

\bibitem[Deng et~al.(2023)Deng, Wang, Hartquist, and Zhao]{deng_nonlocal_2023}
Bolei Deng, Shu Wang, Chase Hartquist, and Xuanhe Zhao.
\newblock Nonlocal {{Intrinsic Fracture Energy}} of {{Polymerlike Networks}}.
\newblock \emph{Phys. Rev. Lett.}, 131\penalty0 (22):\penalty0 228102, December
  2023.
\newblock \doi{10.1103/PhysRevLett.131.228102}.

\bibitem[Hartquist et~al.(2024)Hartquist, Wang, Deng, Beech, Craig, Olsen,
  Rubinstein, and Zhao]{hartquist_fracture_2024}
Chase~M. Hartquist, Shu Wang, Bolei Deng, Haley~K. Beech, Stephen~L. Craig,
  Bradley~D. Olsen, Michael Rubinstein, and Xuanhe Zhao.
\newblock Fracture of polymer-like networks with hybrid bond strengths.
\newblock \emph{Journal of the Mechanics and Physics of Solids}, page 105931,
  November 2024.
\newblock ISSN 0022-5096.
\newblock \doi{10.1016/j.jmps.2024.105931}.

\bibitem[Broedersz et~al.(2011)Broedersz, Mao, Lubensky, and
  MacKintosh]{broedersz_criticality_2011}
Chase~P. Broedersz, Xiaoming Mao, Tom~C. Lubensky, and Frederick~C. MacKintosh.
\newblock Criticality and isostaticity in fibre networks.
\newblock \emph{Nature Phys}, 7\penalty0 (12):\penalty0 983--988, December
  2011.
\newblock ISSN 1745-2481.
\newblock \doi{10.1038/nphys2127}.

\bibitem[Tauber et~al.(2022)Tauber, {van der Gucht}, and
  Dussi]{tauber_stretchy_2022}
Justin Tauber, Jasper {van der Gucht}, and Simone Dussi.
\newblock Stretchy and disordered: {{Toward}} understanding fracture in soft
  network materials via mesoscopic computer simulations.
\newblock \emph{The Journal of Chemical Physics}, 156\penalty0 (16):\penalty0
  160901, April 2022.
\newblock ISSN 0021-9606.
\newblock \doi{10.1063/5.0081316}.

\bibitem[Tankasala and Fleck(2020)]{tankasala_crack_2020}
Harika~C. Tankasala and Norman~A. Fleck.
\newblock The crack growth resistance of an elastoplastic lattice.
\newblock \emph{International Journal of Solids and Structures},
  188--189:\penalty0 233--243, April 2020.
\newblock ISSN 0020-7683.
\newblock \doi{10.1016/j.ijsolstr.2019.10.007}.

\bibitem[de~Waal et~al.(2024)de~Waal, Chouzouris, and
  Dias]{waal_architecting_2024}
Leo de~Waal, Matthaios Chouzouris, and Marcelo~A. Dias.
\newblock Architecting mechanisms of damage in topological metamaterial,
  October 2024.

\bibitem[Curtin and Scher(1990)]{curtin_brittle_1990}
W.~A. Curtin and H.~Scher.
\newblock Brittle fracture in disordered materials: {{A}} spring network model.
\newblock \emph{Journal of Materials Research}, 5\penalty0 (3):\penalty0
  535--553, March 1990.
\newblock ISSN 2044-5326.
\newblock \doi{10.1557/JMR.1990.0535}.

\bibitem[Zhang et~al.(2017)Zhang, Rocklin, Sander, and Mao]{zhang_fiber_2017}
Leyou Zhang, D.~Zeb Rocklin, Leonard~M. Sander, and Xiaoming Mao.
\newblock Fiber networks below the isostatic point: {{Fracture}} without stress
  concentration.
\newblock \emph{Phys. Rev. Mater.}, 1\penalty0 (5):\penalty0 052602, October
  2017.
\newblock \doi{10.1103/PhysRevMaterials.1.052602}.

\bibitem[Rivlin and Thomas(1953)]{rivlin_rupture_1953}
R.~S. Rivlin and A.~G. Thomas.
\newblock Rupture of rubber. {{I}}. {{Characteristic}} energy for tearing.
\newblock \emph{Journal of Polymer Science}, 10\penalty0 (3):\penalty0
  291--318, 1953.
\newblock ISSN 1542-6238.
\newblock \doi{10.1002/pol.1953.120100303}.

\bibitem[Kermode et~al.(2008)Kermode, Albaret, Sherman, Bernstein, Gumbsch,
  Payne, Cs{\'a}nyi, and De~Vita]{kermode_lowspeed_2008}
J.~R. Kermode, T.~Albaret, D.~Sherman, N.~Bernstein, P.~Gumbsch, M.~C. Payne,
  G.~Cs{\'a}nyi, and A.~De~Vita.
\newblock Low-speed fracture instabilities in a brittle crystal.
\newblock \emph{Nature}, 455\penalty0 (7217):\penalty0 1224--1227, October
  2008.
\newblock ISSN 1476-4687.
\newblock \doi{10.1038/nature07297}.

\bibitem[Long and Hui(2016)]{long_fracture_2016}
Rong Long and Chung-Yuen Hui.
\newblock Fracture toughness of hydrogels: Measurement and interpretation.
\newblock \emph{Soft Matter}, 12\penalty0 (39):\penalty0 8069--8086, October
  2016.
\newblock ISSN 1744-6848.
\newblock \doi{10.1039/C6SM01694D}.

\bibitem[Bitzek et~al.(2006)Bitzek, Koskinen, G{\"a}hler, Moseler, and
  Gumbsch]{bitzek_structural_2006}
Erik Bitzek, Pekka Koskinen, Franz G{\"a}hler, Michael Moseler, and Peter
  Gumbsch.
\newblock Structural {{Relaxation Made Simple}}.
\newblock \emph{Phys. Rev. Lett.}, 97\penalty0 (17):\penalty0 170201, October
  2006.
\newblock \doi{10.1103/PhysRevLett.97.170201}.

\bibitem[Gu{\'e}nol{\'e} et~al.(2020)Gu{\'e}nol{\'e}, N{\"o}hring, Vaid,
  Houll{\'e}, Xie, Prakash, and Bitzek]{guenole_assessment_2020}
Julien Gu{\'e}nol{\'e}, Wolfram~G. N{\"o}hring, Aviral Vaid, Fr{\'e}d{\'e}ric
  Houll{\'e}, Zhuocheng Xie, Aruna Prakash, and Erik Bitzek.
\newblock Assessment and optimization of the fast inertial relaxation engine
  (fire) for energy minimization in atomistic simulations and its
  implementation in lammps.
\newblock \emph{Computational Materials Science}, 175:\penalty0 109584, April
  2020.
\newblock ISSN 0927-0256.
\newblock \doi{10.1016/j.commatsci.2020.109584}.

\bibitem[Thompson et~al.(2022)Thompson, Aktulga, Berger, Bolintineanu, Brown,
  Crozier, {in 't Veld}, Kohlmeyer, Moore, Nguyen, Shan, Stevens, Tranchida,
  Trott, and Plimpton]{thompson_lammps_2022}
Aidan~P. Thompson, H.~Metin Aktulga, Richard Berger, Dan~S. Bolintineanu,
  W.~Michael Brown, Paul~S. Crozier, Pieter~J. {in 't Veld}, Axel Kohlmeyer,
  Stan~G. Moore, Trung~Dac Nguyen, Ray Shan, Mark~J. Stevens, Julien Tranchida,
  Christian Trott, and Steven~J. Plimpton.
\newblock {{LAMMPS}} - a flexible simulation tool for particle-based materials
  modeling at the atomic, meso, and continuum scales.
\newblock \emph{Computer Physics Communications}, 271:\penalty0 108171,
  February 2022.
\newblock ISSN 0010-4655.
\newblock \doi{10.1016/j.cpc.2021.108171}.

\bibitem[Dussi et~al.(2020)Dussi, Tauber, and {van der
  Gucht}]{dussi_athermal_2020}
Simone Dussi, Justin Tauber, and Jasper {van der Gucht}.
\newblock Athermal {{Fracture}} of {{Elastic Networks}}: {{How Rigidity
  Challenges}} the {{Unavoidable Size-Induced Brittleness}}.
\newblock \emph{Phys. Rev. Lett.}, 124\penalty0 (1):\penalty0 018002, January
  2020.
\newblock \doi{10.1103/PhysRevLett.124.018002}.

\bibitem[Charles and Hild(2002)]{charles_crack_2002}
Yann Charles and Fran{\c c}ois Hild.
\newblock On crack arrest in ceramic / metal assemblies.
\newblock \emph{International Journal of Fracture}, 115\penalty0 (3):\penalty0
  251--272, June 2002.
\newblock ISSN 1573-2673.
\newblock \doi{10.1023/A:1016399912031}.

\bibitem[Abid et~al.(2019)Abid, Pro, and Barthelat]{abid_fracture_2019}
Najmul Abid, J~William Pro, and Francois Barthelat.
\newblock Fracture mechanics of nacre-like materials using discrete-element
  models: {{Effects}} of microstructure, interfaces and randomness.
\newblock \emph{Journal of the Mechanics and Physics of Solids}, 124:\penalty0
  350--365, March 2019.
\newblock ISSN 0022-5096.
\newblock \doi{10.1016/j.jmps.2018.10.012}.

\bibitem[Persson and Tosatti(2001)]{persson_effect_2001}
Bo~Nils~Johan Persson and Erio Tosatti.
\newblock The effect of surface roughness on the adhesion of elastic solids.
\newblock \emph{J. Chem. Phys.}, 115\penalty0 (12):\penalty0 5597, 2001.
\newblock ISSN 0021-9606.
\newblock \doi{10.1063/1.1398300}.

\bibitem[Evans and Faber(1981)]{evans_toughening_1981}
A.~G. Evans and K.~T. Faber.
\newblock Toughening of {{Ceramics}} by {{Circumferential Microcracking}}.
\newblock \emph{Journal of the American Ceramic Society}, 64\penalty0
  (7):\penalty0 394--398, 1981.
\newblock ISSN 1551-2916.
\newblock \doi{10.1111/j.1151-2916.1981.tb09877.x}.

\bibitem[Kreher and Pompe(1981)]{kreher_increased_1981}
W.~Kreher and W.~Pompe.
\newblock Increased fracture toughness of ceramics by energy-dissipative
  mechanisms.
\newblock \emph{J Mater Sci}, 16\penalty0 (3):\penalty0 694--706, March 1981.
\newblock ISSN 1573-4803.
\newblock \doi{10.1007/BF02402787}.

\bibitem[Evans and Cannon(1986)]{evans_toughening_1986}
A.~G. Evans and R.~M. Cannon.
\newblock Toughening of brittle solids by martensitic transformations.
\newblock \emph{Acta Metall.; (United States)}, 34:5, May 1986.
\newblock \doi{10.1016/0001-6160(86)90052-0}.

\bibitem[Pastewka et~al.(2012)Pastewka, Sharp, and
  Robbins]{pastewka_seamless_2012}
Lars Pastewka, Tristan~A Sharp, and Mark~O Robbins.
\newblock Seamless elastic boundaries for atomistic calculations.
\newblock \emph{Phys. Rev. B}, 86:\penalty0 075459, 2012.
\newblock \doi{10.1103/PhysRevB.86.075459}.

\bibitem[Ma et~al.(2019)Ma, Hajarolasvadi, Albertini, Kammer, and
  Elbanna]{ma_hybrid_2019}
Xiao Ma, Setare Hajarolasvadi, Gabriele Albertini, David~S. Kammer, and
  Ahmed~E. Elbanna.
\newblock A hybrid finite element-spectral boundary integral approach:
  {{Applications}} to dynamic rupture modeling in unbounded domains.
\newblock \emph{Num Anal Meth Geomechanics}, 43\penalty0 (1):\penalty0
  317--338, January 2019.
\newblock ISSN 0363-9061, 1096-9853.
\newblock \doi{10.1002/nag.2865}.

\bibitem[Sinclair(1975)]{sinclair_influence_1975}
J.~E. Sinclair.
\newblock The influence of the interatomic force law and of kinks on the
  propagation of brittle cracks.
\newblock \emph{The Philosophical Magazine: A Journal of Theoretical
  Experimental and Applied Physics}, 31\penalty0 (3):\penalty0 647--671, March
  1975.
\newblock ISSN 0031-8086.
\newblock \doi{10.1080/14786437508226544}.

\bibitem[Hartquist et~al.(2025)Hartquist, Wang, Cui, Matusik, Deng, and
  Zhao]{hartquist_scaling_2025}
Chase Hartquist, Shu Wang, Qiaodong Cui, Wojciech Matusik, Bolei Deng, and
  Xuanhe Zhao.
\newblock Scaling {{Law}} for {{Intrinsic Fracture Energy}} of {{Diverse
  Stretchable Networks}}.
\newblock \emph{Phys. Rev. X}, 15\penalty0 (1):\penalty0 011002, January 2025.
\newblock \doi{10.1103/PhysRevX.15.011002}.

\end{thebibliography}
\end{document}